\documentclass[11pt, twocolumn]{article} 
\usepackage{amsfonts, amssymb, amsmath} 

\usepackage{graphicx}
\usepackage[url=false, eprint=false, doi=false, sorting=none]{biblatex}
\usepackage{xcolor}
\usepackage{multicol}
\usepackage{wrapfig}
\usepackage{ragged2e}
\DeclareMathOperator{\sinc}{sinc}
\bibliography{refs}
\usepackage[margin=0.75in, footskip=0.375in]{geometry} 

\newcommand{\eqn}[0]{\begin{array}{rcl}}
\newcommand{\eqnend}[0]{\end{array}}

\begin{document}
\title{Dynamic distortion from discretized coupling feedback in soft-spin Ising machines}
\author{Victor H. González and Natalia G. Berloff\\
{\it Department of Applied Mathematics and Theoretical Physics}, \\ {\it University of Cambridge, Cambridge CB3 0WA, United Kingdom}}

\date{\today}
\twocolumn[\normalsize
  \begin{@twocolumnfalse}
    \maketitle
        \begin{abstract}
        Although Ising machines (IMs) have grown in popularity as physics-based accelerators for combinatorial optimization, most literature focuses on empirically obtained benchmarks with low universality. We propose dynamic feedback distortion as a universal, platform-agnostic metric for the effect of coupling field discretization in time-multiplexed IMs. Its closed form consists of a static offset and a harmonic series: a mid-tread coupling element (or a half-step bias) removes the former, and injected noise (dither) attenuates the latter. In a simulated optoelectronic IM, 2-bit mid-tread feedback with noise below half a quantization step solves a 50 spin M\"obius ladder in up to 99\% of runs. In a coherent IM, dithered 2-bit mid-tread feedback outperforms 4-bit mid-rise feedback on a M\"obius ladder and matches 4-bit feedback on 2D tiling problems. Triangular dither performs best when the spatial-frequency content of the ground state is unknown. These results show that reducing the static and harmonic components of the network distortion greatly increases the success rate of low-resolution quantized IMs.
\end{abstract}
  \end{@twocolumnfalse}
]
\maketitle

\section*{Introduction}

    Increasing difficulty in miniaturizing semiconductor-based transistors paired with the high power consumption and information bottlenecks of conventional computers have spurred the development of physics-based unconventional computing accelerators to more efficiently solve combinatorial optimization problems (COPs)~\cite{mohseni2022ising, Chowdhury2023-full-stack-p-bits, Gonzalez2024-spintronic-devices-next-generation-computation, Aadit2022-sparsification-fpga, Zhang2024IMReview} as well as meet the growing demands of the current machine learning boom. Ising machines (IMs) are a type of neuromorphic low-power in-memory computer that relies on network relaxation to solve quadratic unconstrained binary optimization (QUBO) problems. At its core, an IM is an embedding of an Ising Hamiltonian into a nonlinear physical network:
    \begin{equation}
        \mathcal{H} = -\frac{1}{2}\sum_{i=1}^N \sum_{j=1}^N J_{ij} s_i s_j 
    \end{equation}
    where the network couplings ($J_{ij}$) and biases ($h_i$) are given by the instance of the quadratic problem to solve, and the system evolves towards an equilibrium configuration $\{s_i\}$ of the binary spin variables $s_i\in \{-1, 1\}$ which minimizes $\mathcal{H}$. Since there exist many mappings between NP-hard problems of industrial and commercial interest and the Ising Hamiltonian~\cite{Barahona1982, Lucas2014IsingFormulations, Barahone1988-circuit-design, babej2018-protein-folding-quantum-annealer}, IMs have been constructed using conventional CMOS technology~\cite{yamaoka2015-20k-spins, cai2020-memristor-hopfield-networks}, oscillator arrays~\cite{Dutta2021VO2-IMs, Borders2019-integer-factorization-mtjs, Chowdhury2023-full-stack-p-bits}, photonic arrays~\cite{Pierangeli2019-photonic-annealer, Pierangeli2020-noise-enhanced-IM}, laser cavities~\cite{Nixon2013CoupledLasers}, parametric optical oscillators~\cite{Haribara2016CIMPerformanceEvalDelayLines, Inagaki2016-Large-scale-IM}, propagating solid state waves~\cite{Mahbood2016-phononicIM, Litvinenko-50spinSAWIM, vadde2026-2048spinbulkacousticwave, ovcharov2024-numerical-model-tm-IM}, opto-electronic circuits~\cite{Sevenants2025-requirements-bit-resolution,Böhm2019PoorMansCIM}, superconducting junctions~\cite{Johnson2011QuantumAnnealer, venturelli2016-job-scheduling-quantum-annealer, albash2018-d-wave-computer}, and even polaritonic condensates~\cite{Berloff2017PolaritonicIMs}. Many of these accelerators rely on digital feedback for coupling, which has unintentional distortion effects of the energy landscapes that these machines explore.
    \begin{figure*}[ht]
        \centering
        \includegraphics[width=\linewidth]{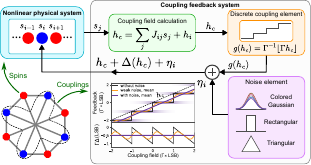}
        \caption{\textbf{Schematic of an IM with distorted coupling feedback.} The network (or graph) associated with a COP instance is realized using two systems. The nonlinear physical system, in light blue, is used for storing and binarizing the spins (nodes) while the coupling feedback system, in grey, constantly measures the spins and updates them according to their couplings (edges). The discrete coupling element enforces the floor function $\lfloor h_c\rfloor$ on the coupling field. The inset shows how white noise increases the noise floor of $\lfloor h_c\rfloor$, distorting the feedback.}
        \label{fig:distorted-IM}
    \end{figure*}
    In a soft-spin IM, shown in Fig.\ref{fig:distorted-IM}, we initially relax the binary constraint of the Ising Hamiltonian and allow spins to have variable amplitude~\cite{Kalinin2018-gain-dissipative-systems}, introducing bifurcation through the inherent nonlinearities of the physical system or via external means such as second harmonic injection locking. The time evolution of soft-spin Ising can be modelled by stochastic differential equations (SDEs) of the form:
    \begin{equation}
        \begin{split}
            \frac{ds_i}{dt} &= f(s_i, t) - g\left(\frac{\partial \mathcal{H}}{\partial s_i}\right)+\zeta(t) \\
            &=f(s_i, t) + g\left(\sum_j J_{ij} s_j \right)+\zeta(t)
        \end{split}
    \end{equation}
    where function $f(s_i, t)$ determines the local dynamics of individual spins, $g\left(\sum_j J_{ij} s_j +h_i\right)$ is the generalized force from coupled spins $s_j$ and $\zeta$ is thermal noise. Although it is often assumed that function $g$ is a linear map of the coupling field $h_c=\sum_j J_{ij} s_j +h_i$ (thus requiring a linear calibration of the feedback by amplification and/or offsetting), this is not the case for many realizations. In coherent~\cite{Honjo2021SciAdv100kCIM, McMahon2016Sci100CIM, Inagaki2016Sci2000CIM, takata16bitCIMdelaylines2016}, solid-state-wave~\cite{litvinenko2023spinwave, Litvinenko-50spinSAWIM,vadde2026-2048spinbulkacousticwave}, and optoelectronic IMs~\cite{Sevenants2025-requirements-bit-resolution}, $h_c$ is calculated digitally with high precision (usually using a field-programmable gate-array or FPGA) and then injected with a discrete coupling element (often an analogue digital converter, or ADC) with limited bitwidth. This digitized feedback introduces an instantaneous field distortion $\Delta(h_c)$, which results in SDEs of the form: 
    \begin{equation}
            \frac{ds_i}{dt} =f(s_i, t) + h_c + \Delta (h_c)+\eta_i(s_i, t)
            \label{eq:distorted_feedback}
    \end{equation}
    where temporal noise $\zeta(t)$ has been replaced by individual spatiotemporal noise $\eta_i(s_i, t)$. While the form of $f(s_i, t)$ depends on each physical realization, the distortion is platform agnostic and its behaviour can be estimated directly from the network couplings and hardware characteristics, namely the bitwidth of the feedback and the amplitude and standard deviation of the noise. From the analytical form of $\Delta(h_c)$, we find a simple distortion mitigation strategy, based on widely used dithering methods, that can help experimental IM researchers maximize the success rate of their setups, regardless of platform. Furthermore, we study the effect of noise with different probability distributions over the distortion and success rate of IMs, and find the triangular white noise is a sensible choice when not much is known about the ground state of the Ising Hamiltonian, while coloured Gaussian noise can be used to leverage the structure of the spin glasses with periodic domain boundaries. We validate our analysis by probing a wide range of problems of increasing complexity and show that distortion mitigation plays an essential role in IMs with rougher energy landscapes.

\subsection*{Distortion of the discretized feedback and mitigation strategy}
    An analytical expression for the distortion $\Delta(h_c)$ can be derived from the network of an IM's problem instance and its coupling feedback system, both shown in Fig.\ref{fig:distorted-IM}. Assuming soft-spins ($|s_i|\leq 1$), the maximum theoretical coupling field given by coupling coefficients $J_{ij}$ is:
    \begin{equation}
        \Phi = \max\limits_{i} \sum_{j=1}^N|J_{ij}|
        \label{eq:max_feedback}
    \end{equation}
    Consider a discrete coupling element, such as an ADC, with a signed bitwidth $b$ and range $[-2^{b}, 2^{b}-1]$. Since it can only represent a finite number of values, we should scale the feedback to represent $\Phi$ with the maximum discrete value and interpolate for smaller $h_c$. Using the whole dynamic range of an ideal mid-rise ADC, a feedback that preserves the ratio between coupling feedback values is given by:
    \begin{equation}
        g(h_c)= \frac{1}{\Gamma}\lfloor\Gamma h_c\rfloor = h_c + \Delta(h_c)
        \label{eq:discrete_feedback}
    \end{equation}
    where the normalization coefficient $\Gamma =\frac{2^b-1}{\Phi}$ can be applied by the $h_c$ calculator while its inverse can be applied by means of an analogue amplifier or attenuator. From Eq.\ref{eq:discrete_feedback} and the Fourier series form of the floor function, we can find an explicit form for $\Delta(h_c)$:
    \begin{equation}
        \Delta(h_c) = -\frac{1}{2\Gamma}+\frac{1}{\pi\Gamma}\sum_{k=1}^\infty\frac{\sin\left(2\pi k \Gamma h_c\right)}{k}
        \label{eq:distortion}
    \end{equation}
    We can directly see the form of the feedback and the distortion in the inset of Fig.~\ref{fig:distorted-IM}, a floor and a sawtooth, respectively. From its analytic expression, we can recognize two features that have been identified empirically by previous studies~\cite{takata16bitCIMdelaylines2016, Sevenants2025-requirements-bit-resolution}: first, that the distortion is inversely proportional to the exponential of the number of bits, meaning that a one or two increase in bitwidth can make an IM much better; and second, that the distortion increases linearly with the maximum theoretical feedback, which means that dense networks or networks with high dynamical ranges between their coupling coefficients require more bits of precision to accurately represent a given Hamiltonian.

    Although mitigating $\Delta(h_c)$ should be a priority in any IM setup, not all IM realizations are flexible enough to introduce elements with larger bitwidth. The static distortion (first term of Eq.\ref{eq:distortion}) can be easily compensated by adding a global biasing field either via the coupling calculator or analogue hardware~\cite{Perdomo-Ortiz2016CorrectionBiasesQuantumAnnealers, Kalinin2018ExternalFields, Gonzalez2023GlobalZTSWIM}. Choosing an appropriate ADC is a simple way to compensate this global distortion since mid-tread ADCs have a built-in half step offset which cancels the first term in the equation.
    
    To reduce the second term, we propose using dither, a well-known technique in signal processing, to reduce the distortion of a quantized signal by the injection of noise~\cite{Lipshitz1992Dither}. This dynamic distortion can be understood as a source of uniform, input dependent, quantization error with amplitude of 1 least significant bit (LSB). Adding zero-mean noise to each individual spin channel helps randomize this error, decorrelating it from the input and turning it into another source of independent noise. This results in the dynamic distortion averaging out (as show in the bottom plot of the inset) and the linearization of the feedback (purple line in the top plot of the inset), at the cost of raising the noise floor of the signal~\cite{Gray1993-ditheredQuantizers,Wannamaker2000-nonsubtractive-dither, Bohm2022-noise-injected-IMs, Pierangeli2020-noise-enhanced-IM}. 
    
    Nonsubtractive dither (as the one showed in Fig.\ref{fig:distorted-IM}), can be modelled by studying the moments of the distribution of the total error introduced by the quantizer. Let a random variable $\nu$ (the dither), independent of the input $h_c$, zero-mean and symmetric, with characteristic function $\Phi_\nu(u)=\mathbb{E}[e^{iu\nu}]$. The total error due to distortion is $\varepsilon(h_c)=\lfloor \Gamma(h_c+\nu)\rfloor/\Gamma -h_c=\nu+\Delta(h_c+\nu)$. The expected value of the total error is:
    
    \begin{equation}
    \mathbb{E}[\varepsilon\,|\,h_c]\;=\;-\frac{1}{2\Gamma}\;+\;\frac{1}{\pi\Gamma}\sum_{k=1}^{\infty}\Phi_\nu\!\left(2\pi k\Gamma\right)\frac{\sin(2\pi k\Gamma h_c)}{k}
    \label{eq:expected_error}
    \end{equation}
     The harmonics of the total error are attenuated by the characteristic function of the dither's amplitude distribution evaluated on the reciprocal lattice $u_k=2\pi k\Gamma$ of the floor function. For fixed $b$ and $\Phi$, adding noise with larger amplitude or standard deviation can help reduce distortion. However, adding noise with very large amplitude or standard deviation affects the quality of the solutions. It is possible to make $\mathbb{E}[\varepsilon\,|\,h_c]$ independent of $h_c$, thus exactly linearizing the quantizer output, by setting $\Phi_\nu(2\pi k\Gamma)=0$ for all $k\geq 1$. 
     
     Zero-mean uniform dither of 1 LSB peak-to-peak (also called rectangular probability density function or RPDF) can provide such characteristic function: $\nu_\text{RPDF} \sim U(-1/2\Gamma, 1/2\Gamma)$, with $\Phi_\text{RPDF}(2\pi k \Gamma)=\sinc(\pi k)=0$ for $k\neq 0$~\cite{Schuchman1964-dither-signals-effects-quantization-noise}. RPDF makes the first moment of the quantization error independent of the input, adding it to the noise floor of the circuit with an input-average error power of $1/6\Gamma^2$. Similarly, one can convolve two RPDF signals to construct triangularly distributed dither with two LSB peak-to-peak (TPDF), although it has a higher power error ($1/4\Gamma^2$), TPDF also makes the second moment of the quantization error input-independent, reducing error modulated distortion at the cost of increasing the noise floor of the system~\cite{Lipshitz1992Dither, Wannamaker2000-nonsubtractive-dither}.

     Coloured Gaussian noise has also been used in previous IM studies ~\cite{Sevenants2025-requirements-bit-resolution, Pierangeli2020-noise-enhanced-IM} and can be understood using the dither formulation. For Gaussian dither $\nu_\text{Gaussian}\sim\mathcal{N}(\mu=0, A, \sigma^2)$ with amplitude $A$ and standard deviation $\sigma$, $\Phi_\text{Gaussian}(2\pi k\Gamma)=\exp{(-2\pi^2k^2\Gamma^2A^2\sigma^2)}$. The distortion harmonics are thus attenuated by a Gaussian profile with e-folding index $k_e=1/(\sqrt{2}\pi\Gamma A\sigma)$. For $\Gamma A\sigma\gtrsim 0.2$, the residual mean distortion is dominated by $k=1$ and bounded by $|\mathbb{E}[\varepsilon\,|\,h_c]-q/2|\leq (1/\Gamma\pi)e^{-2\pi^2\Gamma^2A^2\sigma^2}$. As is the case for RPDF, about half an LSB of Gaussian dither ($\Gamma A\sigma =1/2$) can linearize the quantizer, though not make the distortion input-independent. The proposed dither is somewhat different to conventional subtractive dither and has been chosen for its ease of experimental realization. We expand more on why the choice of having recurrent dither still linearizes the quantizer in Appendix B.
     
     As we will see, making the distortion (and its second moment) error-independent is a good rule-of-thumb for solving combinatorial optimization problems, as this not only linearizes the coupling fields, but also turns $\Delta$, a deterministic quantization error, into a stochastic signal, which is more similar to an effective temperature, which has been shown to benefit the relaxation dynamics of Ising machines. In the next section we report how a small amount of noise improves the solution quality as the distortion is suppressed without modifying the evolution of the IM too severely.
\section*{Results}  
\begin{figure*}[ht!]
        \centering
        \includegraphics[width=\linewidth]{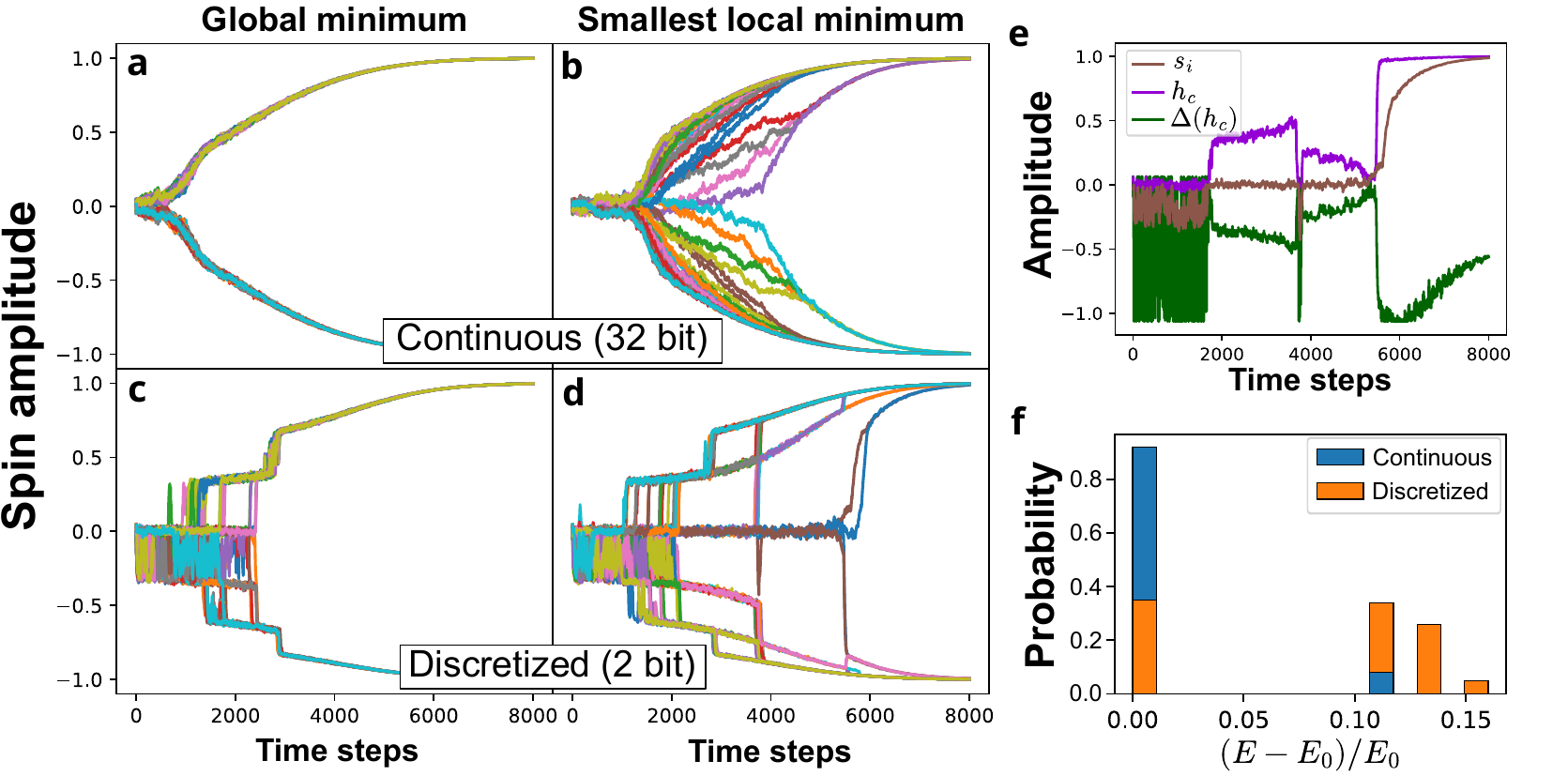}
        \caption{\textbf{The discretization of the field changes the spin dynamics of a 50 spin OEIM solving a Möbius ladder. (a-d)} Spin amplitudes as a function of time with $\kappa=\lambda= A=\sigma=1$. We observe that the discretization of the coupling field induces step-like amplitude growth on the spins, affecting the machine's trajectory and final configuration. \textbf{e.} Spin amplitude, coupling field $h_c$ and distortion $\Delta(h_c)$. For certain spins, the $\Delta(h_c)$ effectively cancels the $h_c$ and their amplitude remains zero for a long interval. \textbf{f.} Probability as a function of normalized energy. Due to distortion, amplitude inhomogeneity promotes local minima and reduces the accuracy of the machine.}
        \label{fig:distorted_binarization}
    \end{figure*}
    \subsection*{Field distortion and spin dynamics}
    
    To study the effects of the distortion and show the versatility of our proposed mitigation strategies, we numerically solved systems of coupled equations representing the time evolution of a soft-spin optoelectronic IM~\cite{Sevenants2025-requirements-bit-resolution} (OEIM) and a coherent Ising machine~\cite{Wang2013-CIM-formulation} (CIM), with increasing graph complexities:
    \begin{equation}
        \frac{ds_i}{dt} = 
        \begin{cases}
            -s_i +\tanh(\kappa s_i +\lambda \lfloor \Gamma h_c\rfloor/\Gamma + \eta_i) & \text{(OEIM)} \\
            (\kappa - 1) s_i - s_i^3 + \lambda \Gamma h_c\rfloor/\Gamma + \eta_i & \text{(CIM)}
        \end{cases}
        \label{eq:OEIM}
    \end{equation}
    where $\kappa$ and $\lambda$ are appropriate hyperparameters for each machine, and $\eta_i$ is the spin-indexed noise. Since optimal hyperparameters $\kappa$, $\lambda$ and the noise amplitude have been reported as problem dependent, we will test our distortion mitigation strategies with fixed $\kappa$ and $\lambda$ and sweeping across the amplitude for seven noise distributions: five coloured Gaussians (brown, pink, white, blue and purple), rectangular (RPDF) and triangular (TPDF) without any annealing schedule.  We find that mitigating the distortion shows significant improvements in the success rates of IMs by sweeping changing the noise amplitude alone.
      
    A Möbius ladder is a type of sparse circulant graph commonly used as a benchmark for IM implementations. In this type of graph, the i-th spin in a chain of length $N$ is coupled antiferromagnetically to its nearest neighbours and to the diametrically opposed spin, i.e. $J_{i,i\pm1}=J_{i,i+\frac{N}{2}}=-1$, as shown in Fig.\ref{fig:distorted-IM}. Max-Cut problems of Möbius graphs are a popular benchmark due to the relative ease of solution verification (the ground state configurations of any ladder with N even are known analytically~\cite{Cummins2025ManifoldReduction}) and computational simplicity~\cite{Kalinin2022ComplexityContinuum}. This universality across physical platforms makes them ideal for distortion characterization and mitigation.


    The spin amplitudes for a soft-spin IM solving the Max-Cut problem of a 50 spin Möbius ladder ($\Phi=3$) with continuous (32 bit, $\Gamma=\frac{2^{32-1}}{3}$) and discretized (2 bit, $\Gamma=1$) feedback are shown in Figs.\ref{fig:distorted_binarization}a-d. For both the global and smallest local minima, field discretization has a clear impact. While the continuous field spins evolve smoothly (Figs.\ref{fig:distorted_binarization}a,b) and their amplitudes grow monotonically, the field distortion induced by the discretization produces discrete amplitude growth in (Figs.\ref{fig:distorted_binarization}c,d) until the spins saturate and binarize. The discretization also seems to delay said binarisation (and convergence to minima) considerably.

    Figure \ref{fig:distorted_binarization}e shows the coupling field $h_c$ and its distortion $\Delta(h_c)$ for a spin $s_i$ (brown spin in Fig.\ref{fig:distorted_binarization}d) with large convergence delay. Since the generalized force (feedback) applied to $s_i$ is $h_c+\Delta(h_c)$, if $\Delta(h_c)\approx-h_c$, then the spin will not grow until other spins have saturated and the field is large enough to overcome the distortion. Similarly, the distortion is responsible for the highly noisy off-zero behaviour in both discretized cases as small spin amplitudes lead to the distortion and noise dominating spin amplitudes (Fig.\ref{fig:distorted_binarization}c,d), slowing the machine's convergence.

    Most importantly for COPs, the discretization-induced distortion also reduces the success rate of IMs. It has been reported before that spin amplitude heterogeneity produces suboptimal spin configurations~\cite{Leleu2019-destabilization-local-minima, Inui2022ArtificialZeeman,Gonzalez2023GlobalZTSWIM}, as shown in Figs.\ref{fig:distorted_binarization}b,d. Due to $\Delta(h_c)$ the amplitudes vary asymmetrically, promoting local minima. In the histogram of Fig.\ref{fig:distorted_binarization}f, constructed using 1000 separate runs with randomized initial conditions, we see the large impact that discretization has over solution quality. The probability of finding the ground state $E_0$ falls from 0.92 to 0.35 when the feedback is discretized and local minima are promoted. While high success probabilities have been reported for Möbius ladders in large systems~\cite{McMahon2016Sci100CIM, Inagaki2016Sci2000CIM, Honjo2021SciAdv100kCIM, Litvinenko-50spinSAWIM, Takesue2025-CIM40k-dense}, such results are obtained after performing problem-dependent hyperparameter optimization (usually through grid search~\cite{Sevenants2025-requirements-bit-resolution} or careful tuning~\cite{litvinenko2023spinwave, vadde2026-2048spinbulkacousticwave}) which increases operational overhead. By contrast, the proposed distortion mitigation protocols relies on adding a static correction bias and spin-indexed noise, both of which can be implemented in analogue hardware with minimal operational overhead.

\subsection*{Effects of noise distribution on distortion}
    To mitigate the distortion, we first suppress the constant term of Eq.\ref{eq:distortion} by adding a static bias of $\frac{1}{2\Gamma}$ to the calculated coupling field before discretization, which is equivalent to replacing the mid-rise quantizer of Eq.\ref{eq:discrete_feedback} with a mid-tread one. We then sample and add the noise with the distributions in Fig.\ref{fig:distorted-IM} at every time step and allow the artificial spin state to equilibrate.
    
\begin{figure}[ht!]
        \centering
        \includegraphics[width=\linewidth]{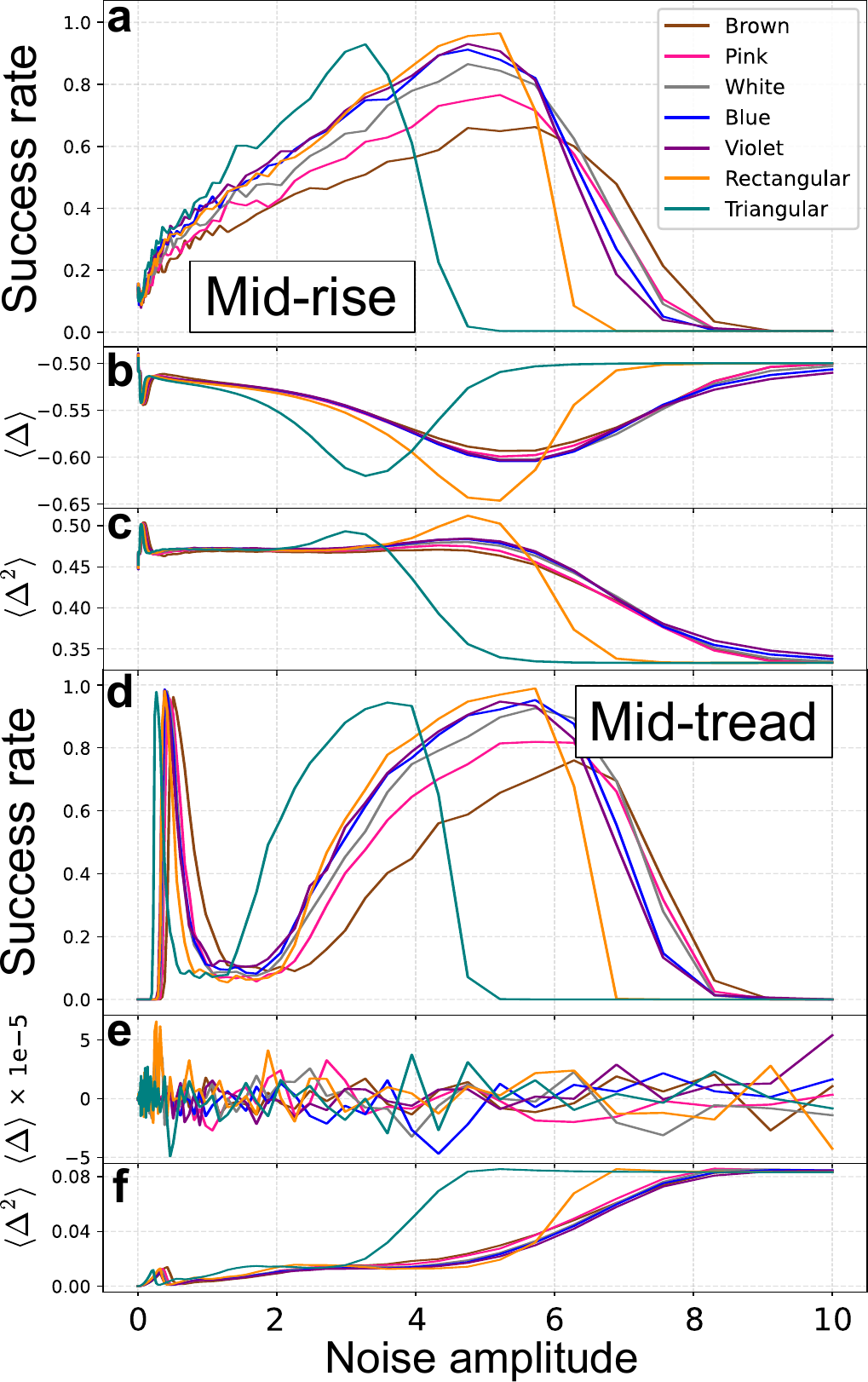}
        \caption{\textbf{50 spin Möbius ladder success rates for 2-bit discretized OEIMs with and without static offset.} Without static offset, a mid-rise ADC \textbf{(a, b, c)} shows success rate improvements only at medium amplitudes while the addition of the $1/2\Gamma$ offset in mid-tread converters \textbf{(d, e, f)} allows dither to linearize smaller feedback values some effectively, enabling very sharp success peaks at low noise amplitudes. In both cases, triangular noise provides the best distortion mitigation. The noise amplitudes are in $h_c$ units.}
        \label{fig:noise-effects}
    \end{figure}

    Figure \ref{fig:noise-effects} shows a comparison between an OEIM solving a 50 spin Möbius ladder with mid-rise and mid-tread discretization with nonsubstractive dither for all seven noise distributions. For the mid-rise ADC of Fig.\ref{fig:noise-effects}a, we observe the noise requires a relatively large amplitude (over 2.5 LSB in this case) to show improvement in convergence in success rate (the ratio between the number of times the machine finds the optimal solution and the total amount of runs). We see that these gradual improvements are correlated to decreases and increases of the total mean distortion $\langle\Delta\rangle$ (Fig.\ref{fig:noise-effects}b) and total second moment of the distortion $\langle\Delta^2\rangle$ (Fig.\ref{fig:noise-effects}c), respectively. This can be understood as a consequence of the converter's so-called dead zone. For this problem, with a 2-bit quantizer and a $1/2$ static bias, small positive amplitudes in the (0,1) interval (the dead zone) do not grow due to their coupling feedback being zero; only growing after the noise pushes them out of this zone. A larger negative mean distortion pushes these small positive amplitudes below zero, giving them a finite negative feedback, and allowing them to grow, raising the success rate of the machine. As expected, when the noise amplitude grows too large, the success rate gradually declines.

    In a mid-tread converter, by contrast, the static offset is already subtracted, allowing the dither to linearize the $h_c$ at much lower noise amplitudes. This is evident in Fig.\ref{fig:noise-effects}d, where all dither distributions show sharp peaks at noise smaller and half one LSB. The first (Fig.\ref{fig:noise-effects}f) and second (Fig.\ref{fig:noise-effects}e) moments of the distortion are also reduced by four and one orders of magnitude, respectively, enabling the IM to more accurately represent the network and linearizing the one LSB wide dead zone at $h_c\approx0$. The reduction of the mean distortion translates into a dynamical relaxation with higher average effective resolution and thus the success rate peaks. These peaks appear in both the OEIM and CIM, as discussed in Appendix A, suggesting this is a feature of the effect of dither on soft-spin IMs rather than a platform specific behaviour.
    
    For both types of ADCs, triangular noise shows the best improvement for the lowest noise amplitude, inline with dither literature~\cite{Wannamaker2000-nonsubtractive-dither, Lipshitz1992Dither, Schuchman1964-dither-signals-effects-quantization-noise}.In signal processing, TPDF is used for signals that are sequentially processed as it is the lowest power ideal dither that linearizes the quantizer without introducing noise modulation. As we will see in the next sections, TPDF is a good dither candidate if the ground state of the Hamiltonian is unknown. If it is known, the colour of the noise can be used to favour relaxation towards the ground state, which also explains the difference in success rates among Gaussian colours. We expand on this point in the section on tunable Möbius ladders below.
    
\section*{Distortion and correction in complex graphs}
    \subsection*{Larger Möbius ladders}   
    \begin{figure*}[ht!]
        \centering
        \includegraphics[width=\linewidth]{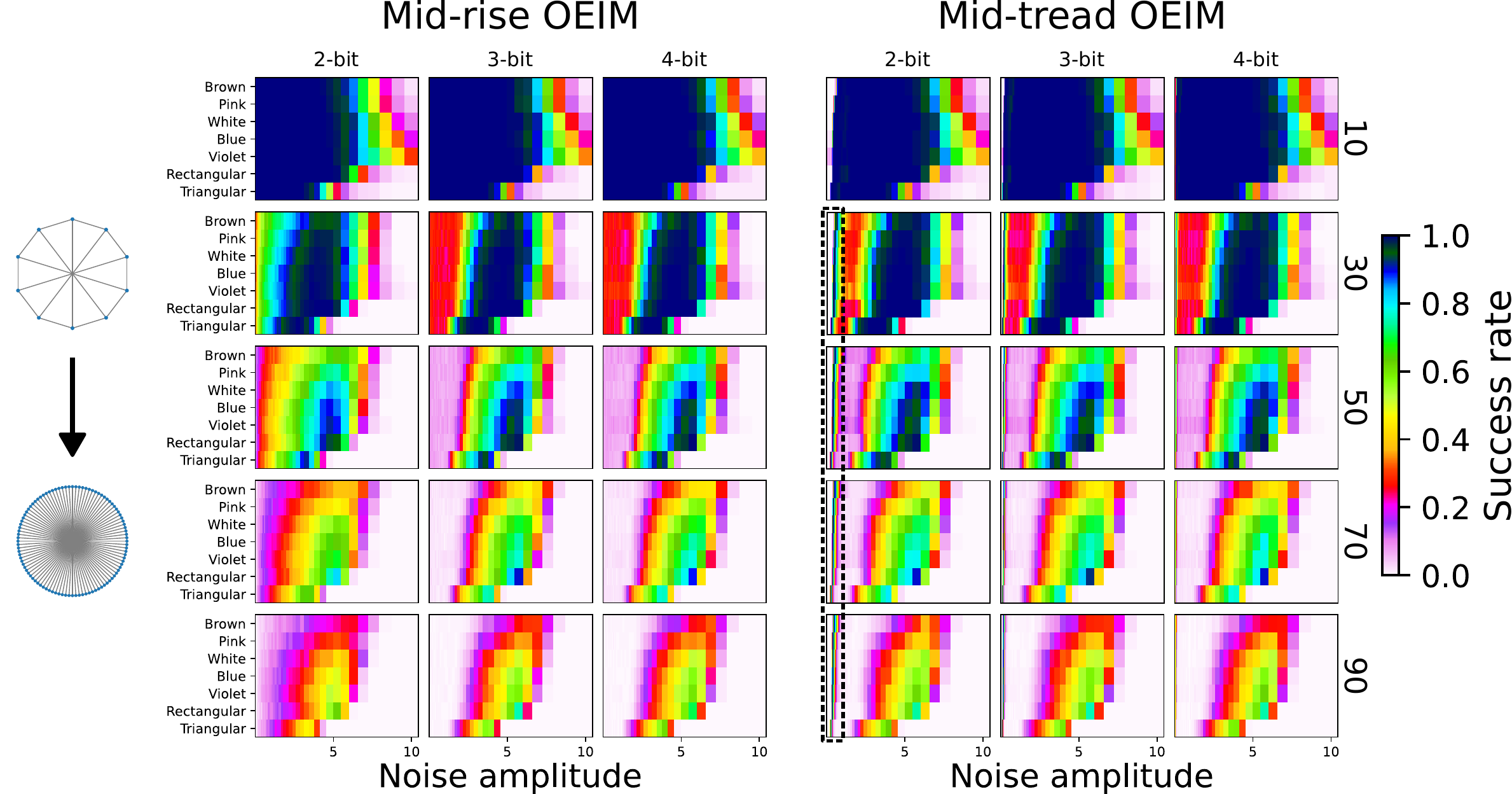}
        \caption{\textbf{Success rates for mid-rise and mid-tread OEIMs for increasingly large Möbius ladders.} Unsurprisingly, with fixed hyperparameters, more complex Möbius ladders have a smaller high success (dark blue-teal) area. Notably, the static offset of mid-tread ADCs allow linearization at lower amplitude, as shown in the dotted black box, producing peaks similar to Fig.~\ref{fig:noise-effects}d. As expected, TPDF performs the best in all cases.}
        \label{fig:mobius-ladders}
    \end{figure*}
    
    We probe how dither scales with problem size by solving the Max-Cut problem for Möbius ladders with 10 to 100 spins with both types of ADC and all noise distributions. Figure \ref{fig:mobius-ladders} shows the success rate as a function of the noise amplitude and the ladder size for multiple feedback bitwidths. For 4 bits of feedback resolution ($\Gamma>5$) and above, there is not much difference between quantizers given since the static and harmonic distortion become exponentially smaller with larger resolution. 

    Two notable features are the ubiquity of TPDF as the best dither option and the presence of low amplitude peaks, such as those observed in the black dotted box of Fig.\ref{fig:noise-effects}, for mid-tread quantizers. An interesting feature of the former, consistent across bitwidths, is that although it requires the smallest noise amplitude, it also shows the narrowest convergence area. This is due to the width of the triangular distribution, since it is 2 LSB peak-to-peak, double that of the others. It is likely that TPDF raises the noise floor faster, which interferes with relaxation towards the ground state. 
    
    The narrowing of the success rate peaks with larger size is inline with an increase in the degeneracy of suboptimal states, suggesting that the noise level exactly linearizing the conditional mean of the error (around 1 LSB) also produces an independent noise floor which acts as an effective temperature and increases the success rate. Since the energy landscape is more rugged for larger graphs, the precision at which this linearization takes place becomes more important as the ladders increase in size. Interestingly, these peaks narrow even more and move to the left for higher resolution quantizers. As described by eq.\ref{eq:distortion}, the distortion in these machines (and the noise necessary to overcome them) will be exponentially smaller. Incidentally, this also makes higher resolution machines more susceptible to noise as dither of a fixed amplitude will encompass a larger amount of bits, acting as a very large effective temperature and hindering their success rates. As shown in Appendix A, these peaks also appear (narrow and disappear) in CIMs, suggesting that they are the results of network distortion operations rather than a specific machine's platform or hyperparameters.

    As a result, we argue that mitigation of the distortion in a network has a larger impact machines with lower resolution. Interestingly enough, as we will see in the other problems are analysed, building IMs with higher resolutions does not seem to have as large of an impact as dithering a low-resolution IMs due to this precise balance that exists when quantization error becomes an effective temperature. For example, for the 90 spin Möbius ladder in Fig.\ref{fig:mobius-ladders}, the narrow low noise peaks for the 2-bit machine reaches a success rate of $\sim1.0$ more effectively than its higher resolution counterparts. This suggests that a low-resolution dithered IM can be better than a higher resolution one with low amplitude uncontrolled noise.

    \subsection*{Tunable Möbius ladders}   
    The hardness of Möbius ladder instances can be tuned continuously by scaling the strength of the cross-circle couplings, $J_{ii+\frac{N}{2}}=-J$ with $J\in [0, 1]$, while keeping the circle couplings at $J_{ii\pm1}=-1$~\cite{Kalinin2022ComplexityContinuum, Cummins2025ManifoldReduction}. Two critical couplings organize the minimization of ladders with even $N/2$~\cite{Cummins2025ManifoldReduction}: at $J_\text{crit}=4/N$, the Ising ground state changes from the fully alternating configuration, which cuts every circle edge, to a partially alternating configuration; and at $J_e=1-\cos(2\pi/N)<J_\text{crit}$, the two largest eigenvalues of the coupling matrix cross and its dominant eigenvector changes symmetry. Gain-based IMs amplify this dominant eigenvector first, so minimization is easy for $J<J_e$, where the eigenvector coincides with the ground state. In the interval $J_e<J<J_\text{crit}$, however, the dominant eigenvector no longer matches the ground state and the machine must escape the states it initially amplifies, making these the most complex instances.
    
    \begin{figure}[ht!]
        \centering
        \includegraphics[width=\linewidth]{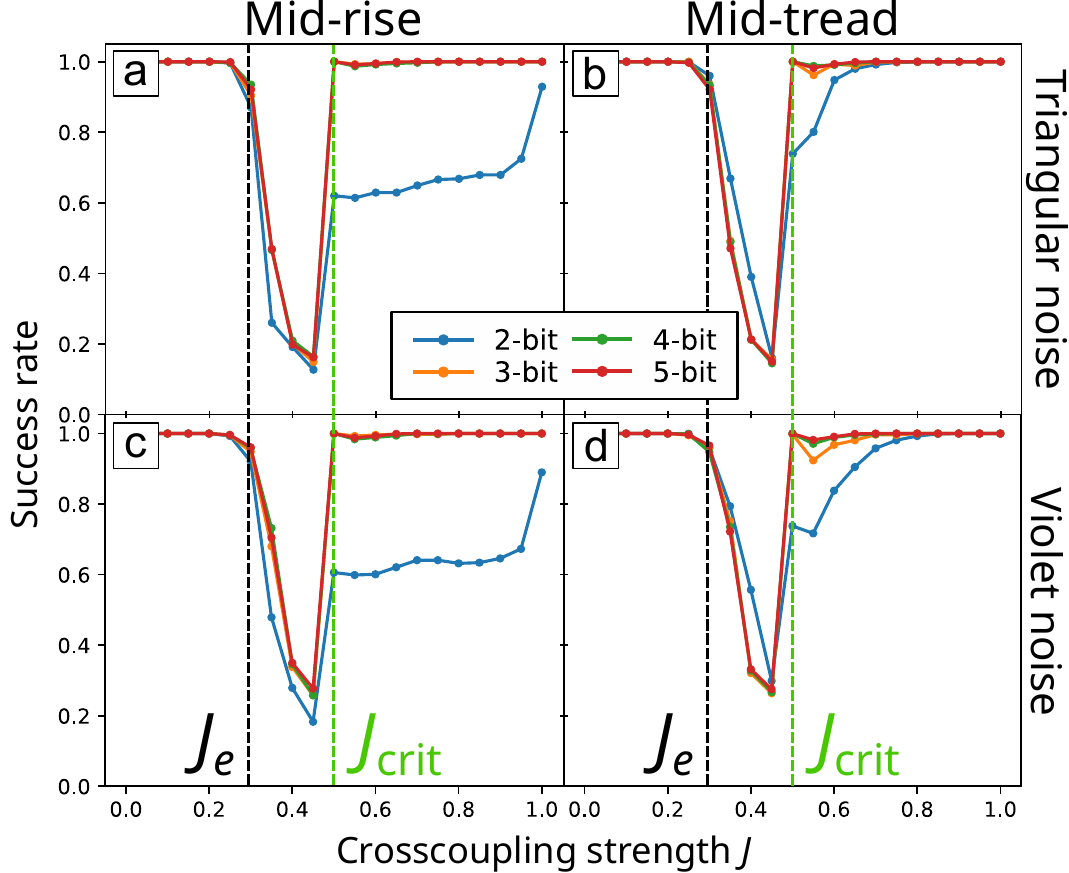}
        \caption{\textbf{Success rate as a function of coupling strength for Möbius ladders with tunable hardness with violet and triangular noise.} As expected, the static distortion of mid-rise low-bit quantizers (\textbf{a,c}) greatly reduces success after the eigenvalue crossing $J_e$ (black dotted lines). By contrast, mid-tread quantizers (\textbf{b,d}) can increase the effectiveness of IMs just by adding a static term. In this case, violet dither is superior to triangular one due to the spin-indexed frequency of the ground state solution.}
        \label{fig:tunable-mobius}
    \end{figure}
    
    Figure \ref{fig:tunable-mobius} shows the success rate as a function of $J$ for ladders with $N=8$ with triangular (\ref{fig:tunable-mobius}a,b) and violet (\ref{fig:tunable-mobius}c,d) dither. Both mid-rise and mid-tread IMs reproduce the expected easy-hard-easy profile: the success rate is near unity below $J_e$, drops in the $J_e$ to $J_\text{crit}$ interval, and grows after $J_\text{crit}$, growing slowly as the new ground state realigns with the dominant eigenvector~\cite{Cummins2025ManifoldReduction}. Similarly to previous sections, subtracting the static distortion is the simplest way to improve the ground state convergence of low resolution IMs. For $J_e<J<$, we observe larger success rates for mid-tread quantizers, and they maintain a consistent advantage for all $J>J_\text{crit}$. An interesting observation is that even beyond the critical coupling, 2-bit mid-rise machines (figs.\ref{fig:tunable-mobius}a,c) are more susceptible to the global $1/2\Gamma$ term, most likely suffering from quantizing effects like amplitude inhomogeneity. Altogether, this shows that reducing $\langle\Delta\rangle$ makes an even better improvement in complex graphs with limited bit precision.

\begin{figure}[ht]
        \centering
        \includegraphics[width=\linewidth]{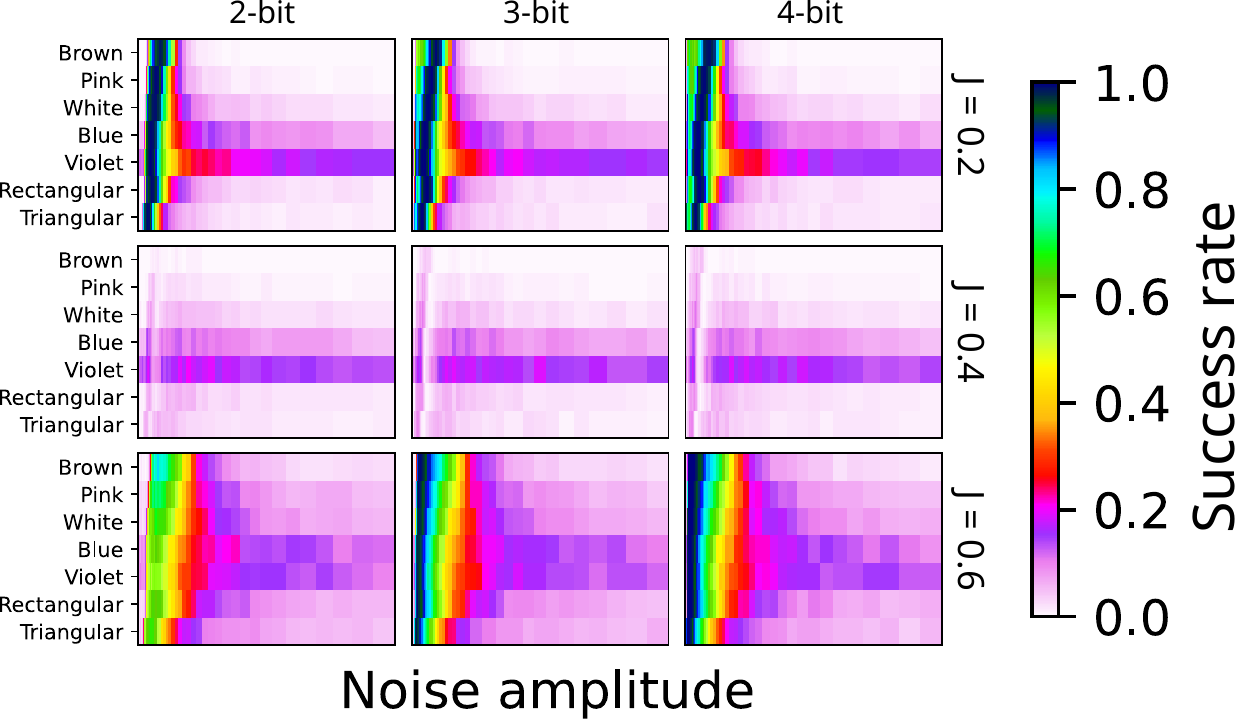}
        \caption{\textbf{Success rates of the tunable Möbius graphs in the region of coupling criticality.} While white triangular noise is a good starting point for any arbitrary Ising problem, coloured dither performs better under specific ground state circumstances. In this case, the alternating spin directions of the circulant graph can be described as having high spin-indexed frequency, similar to violet noise.}
        \label{fig:purple-noise-advantage}
    \end{figure}

    In Fig.\ref{fig:tunable-mobius}, we have chosen to plot violet Gaussian dither alongside TPDF because violet noise works better than all others in the transition region between $J_e$ and $J_\text{crit}$ for all amplitudes and machine configurations. As shown in Fig.\ref{fig:purple-noise-advantage}, blue and violet noise have a broad rate of success rates for all amplitudes, with a clear advantage in the middle of the transition region for J=0.4. This is a consequence of the reciprocal lattice in with the noise is coloured. 
    
    As we mentioned previously, the Gaussian noise is distributed along the spin index number $i$ of the artificial state. Which means that the frequency content of a colour can be expressed as a spatial distribution of the normal modes of the Ising Hamiltonian. For example, a low frequency normal mode will tend to favour a few large sets of adjacent spins with the same orientation (also called domains) with few boundaries between sets (called domain walls). High frequency modes will favour smaller domains and more domain walls. The ground state of a Möbius ladder with N/2 even has the high possible spatial frequencies, both inside and outside of the critical region. In blue and violet Gaussian dither, larger frequencies have a higher power spectral density, favouring these configurations. 
    
    Although appealing, designing a good coloured noise distribution requires knowledge of the solution to the problem. We tested the influence of this indexing in appendix C and found that coloured noise is only an advantage is its spatial frequency content matches that of the ground state of the problem. This could shed more light previous results in quantum annealers~\cite{Liao2022-Quantum-annealing-coloured-noise} and OEIMs~\cite{Sevenants2025-requirements-bit-resolution} and serve as a plausible explanation why noise of a certain colour helps in some problems but not others.

    \subsection*{2D tiling problems}
    
\begin{figure*}[ht!]
        \centering
        \includegraphics[width=\linewidth]{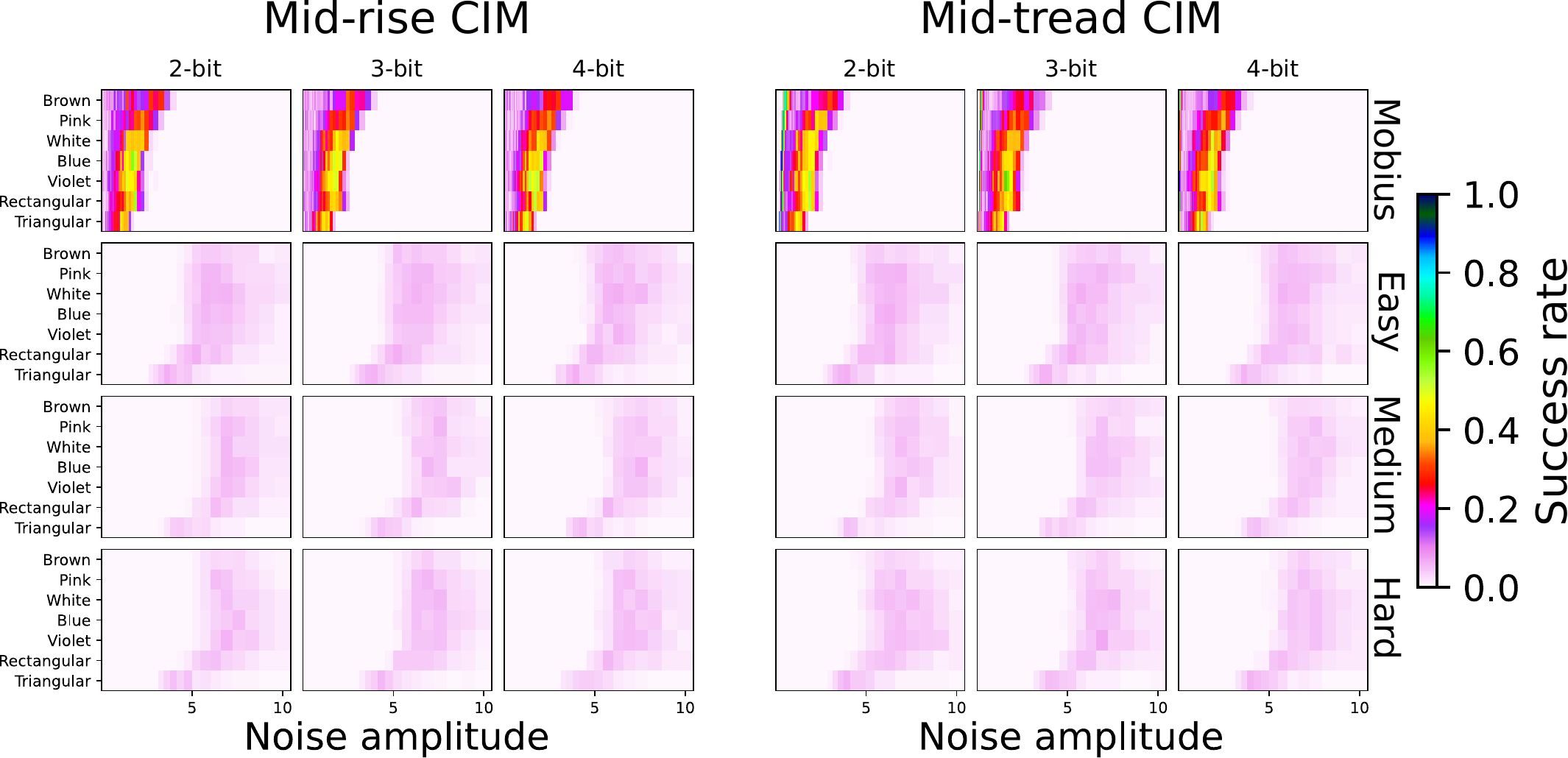}
        \caption{\textbf{Success rates of quantized CIMs for a M\"obius ladder and for 2D tiling problems of increasing hardness.} On the M\"obius ladder, the mid-tread machines show narrow success peaks at amplitudes of half an LSB or less, which move to lower amplitudes as bits are added, while the mid-rise machines peak at lower success rates and larger amplitudes. In the tilings, the success rate is negligible at low amplitudes and peaks at amplitudes that do not depend on the quantizer or its bitwidth, with TPDF peaking at the lowest amplitude in every panel. The noise amplitudes are in $h_c$ units and all graphs have the same ampunt of spins.}
        \label{fig:2D-tiling}
    \end{figure*}

    As a final benchmark, we consider previously studied 2D tiling problems~\cite{Syed2026VectorSpins}. These instances are generated with the tile planting method implemented in the Chook Python library~\cite{perera2021CHook}: a square lattice is tiled with frustrated unit cells drawn from a small set of cell families, and the mixing probabilities of the families tune the computational hardness of the resulting Ising problem while guaranteeing that the planted ground state is known by construction. We solved ten instances of each of the easy, medium and hard classes in OEIMs and CIMs (see Appendix A for the full results), sweeping the amplitude of all seven noise distributions, and used a M\"obius ladder solved with the same machines as a reference. The resulting success rates are shown in Fig.~\ref{fig:2D-tiling}.

    The M\"obius ladder in the top row of Fig.~\ref{fig:2D-tiling} reproduces with the CIM the behaviour observed with the OEIM (Fig.~\ref{fig:noise-effects} and Appendix~{A}). Without noise, the 2- and 3-bit mid-tread machines do not converge: their dead zone maps the small initial fields onto zero, and the amplitudes decay. A small amount of noise lifts this restriction and produces narrow success peaks between 0.80 and 0.96. As bits are added, these peaks move to lower amplitudes in proportion to the quantization step $1/\Gamma$, from $A\approx0.2$--$0.5$ with 2 bits to $A\approx0.1$--$0.2$ with 3 bits and $A\approx0.05$--$0.09$ with 4 bits, which corresponds to a noise standard deviation of about $0.1$~LSB for all seven distributions (see Methods). The mid-rise machines peak at lower success rates, between 0.30 and 0.57, and at larger amplitudes, $A\approx1$--$3$, for every bitwidth; even with 4 bits, they remain below the 2-bit mid-tread machine. Both features follow from the two components of Eq.~\ref{eq:distortion}: the static term is removed by the choice of quantizer, and the harmonic terms by noise whose optimal amplitude is set by the quantization step.
    
    The tilings are harder problems for soft-spin IMs. Their peak success rates, between approximately 0.03 and 0.07, are an order of magnitude lower than those of the M\"obius ladder, and they require much stronger noise: below $A\approx2$, the success rate is practically zero for every noise distribution, quantizer and bitwidth, so that in these problems (without hyperparamenter optimisation) the noise is a condition for convergence rather than a refinement. The success rate rises from $A\approx3$ for TPDF, $A\approx4$--$5$ for RPDF and $A\approx5$--$6$ for the Gaussian colours, peaks at $A\approx4$, $5$--$6$ and $6$--$7$, respectively, and decreases towards $A=10$. In contrast with the M\"obius ladder, neither the position nor the height of these peaks depends systematically on the bitwidth or on the type of quantizer: the 2-bit machines reach the same success rates as the 4-bit ones, and the differences between the mid-rise and mid-tread machines are comparable to the binomial standard error of 1000 runs at these success rates, $\approx0.007$. The optimal noise is therefore set by the energy landscape of the instance rather than by the quantization step. At these amplitudes, the harmonic terms of Eq.~\textcolor{red}{7} are strongly attenuated for both quantizers, and the static term of the mid-rise element, which zero-mean noise cannot remove, has no measurable effect on the success rate. For problems of this kind, a 2-bit coupling element with injected noise performs as well as a 4-bit one, which relaxes the bit-resolution requirements reported for analogue IMs~\cite{Sevenants2025-requirements-bit-resolution}.
    
    The choice of noise distribution has a clearer effect than that of the quantizer. Averaged over the three classes, the three bitwidths and the two quantizers, the peak success rate of TPDF, $\approx0.05$, is the highest of the seven distributions, followed closely by those of white Gaussian noise and RPDF, the two other distributions with a flat spectrum along the spin index. TPDF also reaches its peak at the lowest amplitude in every panel, $A\approx3.6$--$4.3$, against $A\approx4.8$--$6.3$ for RPDF and $A\approx5.2$--$8.3$ for the Gaussian colours. This ordering follows from the width of the distributions rather than from a different optimal noise power: at a given amplitude, the standard deviation of TPDF, $A/\sqrt{6}$, is larger than those of RPDF, $A/\sqrt{12}$, and of the Gaussian colours, $A/4$ (see Methods), and all seven distributions peak at a standard deviation of approximately $1.65$ in units of the coupling field. Among the Gaussian colours, brown noise, which concentrates its power at the lowest spatial frequencies along the spin index, gives the lowest peak success rates in all three classes, roughly 25--35\% below the other distributions, whereas none of the remaining colours is consistently ahead: the best-performing colour changes between classes, bitwidths and quantizers.
    
    This behaviour differs from that of the tunable M\"obius ladders, where violet noise outperforms the other distributions because its spectrum matches the alternating ground state along the spin index (Figs.~\ref{fig:tunable-mobius} and ~\ref{fig:purple-noise-advantage}). Such matching requires prior knowledge of the ground state, which is available for benchmark problems such as the M\"obius ladders but not before an arbitrary problem is solved. In the tilings, no colour has a systematic advantage. Triangular dither does not depend on this information: its spectrum along the spin index is flat, so it does not favour any spatial frequency of the solution, and it is the lowest-power dither that makes both the first and the second moments of the quantization error independent of the input~\cite{Lipshitz1992Dither, Wannamaker2000-nonsubtractive-dither}. Together with its performance on the M\"obius ladders (Figs.~\ref{fig:noise-effects} and~\ref{fig:mobius-ladders}), this makes TPDF the best choice among the distributions studied here, and a good dither candidate whenever no information on the spatial-frequency content of the ground state is available; coloured noise can then be considered as a refinement once that content is known.

\section*{Discussion}

    In this work, we introduced the dynamic feedback distortion $\Delta(h_c)$ as a universal, platform-agnostic metric for the effect of coupling field discretization in time-multiplexed IMs. Starting from the transfer function of a realistic discrete coupling element, we derived a closed form for the distortion (Eq.~\ref{eq:distortion}) with two components: a static offset of half a quantization step, $-1/(2\Gamma)$, and a harmonic series in the coupling field. We showed analytically and numerically how both components can be suppressed: the static one by using a mid-tread coupling element or, equivalently, by adding a bias of $1/(2\Gamma)$ to the coupling field before a mid-rise element, and the harmonic one by injecting zero-mean noise, which attenuates the $k$-th harmonic by the characteristic function of its distribution evaluated at $2\pi k\Gamma$ (Eq.~\ref{eq:expected_error}).

    The two elements differ only in the position of their decision thresholds relative to the output levels, but this difference changes the dynamics of low-resolution machines. A mid-rise element places its thresholds at the integer multiples of the quantization step $1/\Gamma$ and truncates the field to the level below it, whereas a mid-tread element places them halfway between levels and rounds the field to the nearest one. Both map an interval one LSB wide onto zero, $[0,1/\Gamma)$ for the mid-rise element and $|h_c|<1/(2\Gamma)$ for the mid-tread one, but only the former is asymmetric: small negative fields are sent to $-1/\Gamma$, while small positive ones receive no feedback. Moreover, since $\Gamma\Phi=2^b-1$ is an integer, a mid-rise element has a threshold exactly at the largest field, $\Phi$. On the M\"obius ladders, where a saturated spin with all its bonds satisfied experiences a field of $\pm\Phi$, amplitudes that saturate from below therefore lose a whole quantization step of feedback when their field is positive and none when it is negative. This accounts for the mean distortion of $\approx-1/(2\Gamma)$ and the second moment of $\approx1/(2\Gamma^2)$ of the mid-rise OEIM at low noise (Figs.~\ref{fig:noise-effects}b,c), and is consistent with the amplitude heterogeneity of Fig.~\ref{fig:distorted_binarization}. In a mid-tread element, $\pm\Phi$ lie at the centre of their quantization intervals and are reproduced exactly, and $\langle\Delta\rangle$ is reduced by four orders of magnitude (Fig.~\ref{fig:noise-effects}e). Its symmetric dead zone, however, holds the small initial amplitudes at zero: without noise, none of the runs of the 2-bit mid-tread machines converged, in either the OEIM or the CIM (Figs.~\ref{fig:noise-effects}d and~\ref{fig:2D-tiling}). The static correction and the noise are therefore complementary: the former removes the offset and the asymmetry it creates between up and down spins, and the latter linearizes the quantizer around the origin and allows the amplitudes to leave the dead zone earlier.

    From our results, we identify four design principles for time-multiplexed IMs with quantized feedback. First, the coupling element should be mid-tread or, if only a mid-rise element is available, its input should be biased by $1/(2\Gamma)$. This cancels the static term of the distortion at no additional hardware cost. Second, the static correction should be combined with noise, which suppresses the harmonic terms and the dead zone. On the M\"obius ladders, the optimal noise of the mid-tread machines scales with the quantization step, with a standard deviation of about $0.1$~LSB for every distribution and bitwidth, so that a small amount suffices: at amplitudes below half an LSB, the 2-bit mid-tread OEIM solved the 50 spin ladder in 96--99\% of the runs, against at most 35\% for the mid-rise OEIM at the same amplitudes (Figs.~\ref{fig:noise-effects}a,d). In the tilings, stronger noise is needed, and its optimal amplitude is set by the energy landscape rather than by the quantizer. In both cases, the bias and the noise can be implemented with simple analogue hardware and minimal operational overhead~\cite{Lee2025-noise-augmented-p-bits}.
    
    Third, in the absence of information on the ground state, TPDF is the most reliable choice of noise: it linearizes the quantizer at the lowest amplitude on the M\"obius ladders (Figs.~\ref{fig:noise-effects} and~\ref{fig:mobius-ladders}) and gave the highest average peak success rate on the tilings. Coloured Gaussian noise is advantageous only when its spectrum matches the spatial-frequency content of the ground state along the spin index, as violet noise~\cite{Liao2022-Quantum-annealing-coloured-noise} does for the tunable M\"obius ladders inside the $J_e$ to $J_{\mathrm{crit}}$ interval (Figs.~\ref{fig:tunable-mobius} and~\ref{fig:purple-noise-advantage}); when it does not, as for brown noise in the tilings, it lowers the success rate. Fourth, distortion mitigation has the largest impact on low-resolution machines solving complex problems: its advantage grows with the ladder size (Fig.~\ref{fig:mobius-ladders}), it doubles the success rate inside the hard $J_e$ to $J_{\mathrm{crit}}$ interval of the tunable ladders (Fig.~\textcolor{red}{5}), and it allows a 2-bit mid-tread CIM to outperform a 4-bit mid-rise one on the M\"obius ladder, with peak success rates of 0.80--0.92 against at most $\approx0.52$ (Fig.~\ref{fig:2D-tiling}). In the 2D tilings, where the success rate without noise was practically zero, the injected noise allowed the 2-bit machines to reach the same success rates as the 4-bit ones.

    Taken together, these results show that reducing the static and harmonic components of the network distortion greatly increases the success rate of low-resolution quantized IMs. Adding bits reduces both components only by about a factor of two per bit, at the cost of a more complex coupling element, whereas a static bias and a small amount of dither suppress them at the resolution already available: a dithered 2-bit mid-tread machine outperformed a 4-bit mid-rise machine on the M\"obius ladder, and dithered 2-bit machines matched the 4-bit ones on the tilings.

    Our derivation only assumes a time-multiplexed architecture in which the coupling field is computed digitally and injected through an element of finite bitwidth. The framework is therefore independent of the nonlinearity $f(s_i,t)$ of the host system and can be applied to all time-multiplexed IMs, including coherent~\cite{Honjo2021SciAdv100kCIM,McMahon2016Sci100CIM,Inagaki2016Sci2000CIM,takata16bitCIMdelaylines2016}, solid-state-wave~\cite{litvinenko2023spinwave, Litvinenko-50spinSAWIM,vadde2026-2048spinbulkacousticwave} and optoelectronic~\cite{Bohm2021,Sevenants2025-requirements-bit-resolution} realizations, as the similar behaviour of the OEIM and the CIM in our simulations indicates. Because $\Delta(h_c)$ follows from the coupling matrix of the problem and from the bitwidth, the threshold positions and the noise of the coupling element alone, it can be estimated before a machine is built and used to choose the resolution and type of the ADC, as well as the amount and distribution of the noise to inject, for a target class of problems. In this work, the hyperparameters $\kappa$ and $\lambda$ and the noise amplitude were held constant during each run to simplify the assumptions of the model. Future work will combine distortion mitigation with annealing schedules for the noise or the gain to further improve and refine the framework as well as extend it to spatially resolved IMs. Based on our results, distortion mitigation via dithering is a powerful tool which can guide operational protocols to improve IMs with heavily quantized feedback resolution.
    
\section*{Methods}
    Unless stated otherwise, Eqs.~\ref{eq:OEIM} were integrated with $\kappa=0.5$ and $\lambda=1$ for the OEIM and $\kappa=\lambda=0.5$ for the CIM, so that both machines have the same linear self-feedback, $\kappa-1=-0.5$, and the spins only grow through the coupling. Each parameter point consists of 1000 independent runs, with initial amplitudes drawn uniformly from $[-0.05,0.05]$, which are advanced with a fixed time step $\delta t$ as
    \begin{align*}
    s_i &\leftarrow s_i + \delta t\,\bigl[-s_i + \tanh(\kappa s_i + \lambda Q_i + \eta_i)\bigr],\\
    s_i &\leftarrow s_i + \delta t\,\bigl[(\kappa-1)s_i - s_i^3 + \lambda Q_i\bigr] + \sqrt{\delta t}\,\eta_i,
    \end{align*}
    for the OEIM and the CIM, respectively, where $Q_i=Q(h_c)=\lfloor\Gamma(h_c+c)\rfloor/\Gamma$ is the quantized field of spin $i$ and the noise $\eta_i$ is drawn anew at every step. The OEIM is integrated with the explicit Euler method, with $\delta t=0.5$ over a total time $T=1500$ (3000 steps), the noise entering the argument of the hyperbolic tangent together with the quantized feedback. The CIM is integrated with the Euler--Maruyama method, with $\delta t=0.05$ over $T=300$ (6000 steps). The shorter step avoids the instability of the explicit cubic update, $s_i(1-\delta t\,s_i^2)$, which grows without bound once $\delta t\,s_i^2>2$, a condition reached by the largest noise amplitudes at $\delta t=0.5$; the factor $\sqrt{\delta t}$ keeps the noise intensity per unit time independent of the step. In both machines, the noise is added to the quantized feedback, as in Fig.~\ref{fig:distorted-IM}, and reaches the coupling element through the spin amplitudes at the following steps.
    
    The coupling matrix of each instance, either the circulant matrix of a M\"obius ladder or a planted tiling generated with Chook \cite{perera2021CHook}, determines $\Phi$ and $\Gamma=(2^b-1)/\Phi$ through Eqs.~\ref{eq:max_feedback} and~\ref{eq:discrete_feedback}; on the M\"obius ladders ($\Phi=3$), $\Gamma=1$, $7/3$ and $5$ for 2, 3 and 4 bits. At every step, the coupling field $h_c=\sum_j J_{ij}s_j$ (all instances studied here have $h_i=0$) is computed for all spins and runs as a matrix product, and the discrete coupling element is emulated with $c=0$ for the mid-rise element and $c=1/(2\Gamma)$ for the mid-tread one. No clipping is applied to the quantizer output: in the OEIM, $|s_i|\le1$ guarantees $|h_c|\le\Phi$, while in the CIM the amplitudes remain below $\approx(\kappa-1+\lambda\Phi)^{1/2}$ in the absence of noise, which equals unity on the M\"obius ladders with the hyperparameters above. The distortion is evaluated exactly at every step, $\Delta=Q(h_c)-h_c$, rather than from its series (Eq.~\ref{eq:distortion}), and its first and second moments, $\langle\Delta\rangle$ and $\langle\Delta^2\rangle$ in Fig.~\ref{fig:noise-effects}, are averages over time steps, spins and runs. They are expressed in units of the coupling field rather than in LSB, so that machines with different bitwidths can be compared.
    
    The noise amplitude $A$ was swept over 100 values logarithmically spaced between $10^{-3}$ and $10$, in units of the coupling field, with $\sigma=1$. RPDF and TPDF dither are drawn independently for every spin, run and step as $\eta_i=A\sigma(u_i-1/2)$ and $\eta_i=A\sigma(u_i+u_i'-1)$, with $u_i$ and $u_i'$ uniform on $[0,1)$, so that $A$ is the peak-to-peak width of the RPDF and half that of the TPDF; for $\Gamma=1$, an amplitude $A=1$ spans one and two LSB, respectively, as in the dither analysis above. Coloured Gaussian noise is shaped along the spin index: a vector $z$ of independent standard normal variables is drawn at every step and transformed as $\eta=(A\sigma/4)\,Fz$, where $F$ is a precomputed real $N\times N$ matrix that multiplies the discrete Fourier transform of $z$ along $i$ by an amplitude profile proportional to $f^{-\alpha/2}$, normalized to unit root mean square over the non-negative frequencies, and transforms back~\cite{Kasdin1995ColouredNoise}. This yields power spectral densities proportional to $f^{-\alpha}$, with $\alpha=2$, $1$, $0$, $-1$ and $-2$ for brown, pink, white, blue and violet noise, respectively. The zero-frequency component vanishes for every colour except white, so coloured noise has no uniform component across the spins, and the noise is white in time for all distributions. With these definitions, the per-spin standard deviation at amplitude $A$ is $A/\sqrt{6}\approx0.41A$ for TPDF, $A/\sqrt{12}\approx0.29A$ for RPDF and $\approx A/4$ for the Gaussian colours.
    
    At the end of each run, the spins are binarized by their sign and the Ising energy $E$ of the resulting configuration is computed with Eq.~\textcolor{red}{1}. A run is counted as successful when $|E-E_0|\le\rho|E_0|$, where $E_0$ is the ground-state energy, known analytically for the M\"obius ladders~\cite{Cummins2025ManifoldReduction} and by construction for the planted tilings~\cite{perera2021CHook, Syed2026VectorSpins}, and $\rho$ is a relative radius, set to 1\%. For the M\"obius ladders with unit couplings and up to 100 spins, whose energy levels are separated by at least 2 and whose ground-state energies satisfy $|E_0|\le150$, this criterion is equivalent to reaching the exact ground state. The success rate is the fraction of successful runs at each amplitude; the final energy of every run can also be stored, so that success rates for other radii can be evaluated from the same simulations.
    
    The simulations were run in double precision with NumPy on multi-core CPUs in at the Cambridge Service for Data Driven Discovery (CSD3). For every noise distribution and amplitude, the runs are split into batches that are propagated together as matrix operations and distributed over independent worker processes, one per available CPU, each restricted to single-threaded linear algebra. Every batch draws from its own random stream, an SFC64 generator seeded through NumPy's \texttt{SeedSequence} from a global seed and the position of the batch in the sweep, so that, for a given seed and batch size, the results do not depend on the number of processes. The code used for the simulations is available in the supplemental material.
    
\section*{Data Availability Statement}
    The code used in these simulations is included with the supplemental material of this article, along with a README file describing CLI usage, package dependencies and the specific seeds used for reproducibility purposes.
    
\section*{Acknowledgements}
    The authors acknowledge the support from the HORIZON EIC-2022-PATHFINDER CHALLENGES-01 HEISINGBERG Project 101114978. N.G.B. acknowledges the support from  the  EPSRC UK Multidisciplinary Centre for Neuromorphic Computing (grant UKRI982), and Weizmann-UK Make Connection Grant 142568.
\medskip
\printbibliography

@misc{Syed2026VectorSpins,
      title={Soft vector spins with dimensional annealing for combinatorial optimization}, 
      author={Marvin Syed and Richard Zhipeng Wang and Natalia G. Berloff},
      year={2026},
      eprint={2604.01003},
      archivePrefix={arXiv},
      primaryClass={cond-mat.dis-nn},
      url={https://arxiv.org/abs/2604.01003}, 
}

@article{perera2021CHook,
      title={Chook -- A comprehensive suite for generating binary optimization problems with planted solutions}, 
      author={Dilina Perera and Inimfon Akpabio and Firas Hamze and Salvatore Mandra and Nathan Rose and Maliheh Aramon and Helmut G. Katzgraber},
      year={2021},
      journal={arXiv preprint: [quant-ph] 2005.14344},
      url={https://arxiv.org/abs/2005.14344}, 
}

@article{Wang2013-CIM-formulation,
  title = {Coherent Ising machine based on degenerate optical parametric oscillators},
  author = {Wang, Zhe and Marandi, Alireza and Wen, Kai and Byer, Robert L. and Yamamoto, Yoshihisa},
  journal = {Phys. Rev. A},
  volume = {88},
  issue = {6},
  pages = {063853},
  numpages = {9},
  year = {2013},
  month = {Dec},
  publisher = {American Physical Society},
  doi = {10.1103/PhysRevA.88.063853},
  url = {https://link.aps.org/doi/10.1103/PhysRevA.88.063853}
}

@ARTICLE{Schuchman1964-dither-signals-effects-quantization-noise,
  author={Schuchman, L.},
  journal={IEEE Transactions on Communication Technology}, 
  title={Dither Signals and Their Effect on Quantization Noise}, 
  year={1964},
  volume={12},
  number={4},
  pages={162-165},
  doi={10.1109/TCOM.1964.1088973}}

@Article{Sevenants2025-requirements-bit-resolution,
author={Sevenants, Toon
and Van der Sande, Guy
and Verschaffelt, Guy},
title={Requirements on bit resolution in optical Ising machine implementations with analog spin variables},
journal={Communications Physics},
year={2025},
month={Jan},
day={04},
volume={8},
number={1},
pages={11},
issn={2399-3650},
doi={10.1038/s42005-024-01919-9},
}

@ARTICLE{Gray1993-ditheredQuantizers,
  author={Gray, R.M. and Stockham, T.G.},
  journal={IEEE Transactions on Information Theory}, 
  title={Dithered quantizers}, 
  year={1993},
  volume={39},
  number={3},
  pages={805-812},
  doi={10.1109/18.256489}}

@ARTICLE{Wannamaker2000-nonsubtractive-dither,
  author={Wannamaker, R.A. and Lipshitz, S.P. and Vanderkooy, J. and Wright, J.N.},
  journal={IEEE Transactions on Signal Processing}, 
  title={A theory of nonsubtractive dither}, 
  year={2000},
  volume={48},
  number={2},
  pages={499-516},
  doi={10.1109/78.823976}}

@article{Lipshitz1992Dither,
  author = {Lipshitz, Stanley P. and Wannamaker, Robert A. and Vanderkooy, John},
  title = {{Quantization and Dither: A Theoretical Survey}},
  journal = {Journal of the Audio Engineering Society},
  volume = {40},
  number = {5},
  pages = {355--375},
  year = {1992},
  month = may,
  publisher = {Audio Engineering Society},
  url = {https://aes.org/publications/elibrary-page/?id=7047}
}

@article{Pierangeli2020-noise-enhanced-IM,
author = {Pierangeli, Davide and Marcucci, Giulia and Brunner, Daniel and Conti, Claudio},
title = {Noise-enhanced spatial-photonic Ising machine},
journal = {Nanophotonics},
volume = {9},
number = {13},
pages = {4109-4116},
doi = {https://doi.org/10.1515/nanoph-2020-0119},
url = {https://onlinelibrary.wiley.com/doi/abs/10.1515/nanoph-2020-0119},
eprint = {https://onlinelibrary.wiley.com/doi/pdf/10.1515/nanoph-2020-0119},
year = {2018}
}

@article{Liao2022-Quantum-annealing-coloured-noise,
author = {Liao, Zhiqiang and Ma, Kaijie and Sarker, Md Shamim and Tang, Siyi and Yamahara, Hiroyasu and Seki, Munetoshi and Tabata, Hitoshi},
title = {Quantum Analog Annealing of Gain-Dissipative Ising Machine Driven by Colored Gaussian Noise},
journal = {Advanced Theory and Simulations},
volume = {5},
number = {3},
pages = {2100497},
doi = {https://doi.org/10.1002/adts.202100497},
url = {https://advanced.onlinelibrary.wiley.com/doi/abs/10.1002/adts.202100497},
eprint = {https://advanced.onlinelibrary.wiley.com/doi/pdf/10.1002/adts.202100497},
year = {2022}
}

@article{Nixon2013CoupledLasers,
  title = {Observing Geometric Frustration with Thousands of Coupled Lasers},
  author = {Nixon, Micha and Ronen, Eitan and Friesem, Asher A. and Davidson, Nir},
  journal = {Phys. Rev. Lett.},
  volume = {110},
  issue = {18},
  pages = {184102},
  numpages = {5},
  year = {2013},
  month = {May},
  publisher = {American Physical Society},
  doi = {10.1103/PhysRevLett.110.184102},
  url = {https://link.aps.org/doi/10.1103/PhysRevLett.110.184102}
}

@Article{Inagaki2016-Large-scale-IM,
author={Inagaki, Takahiro
and Inaba, Kensuke
and Hamerly, Ryan
and Inoue, Kyo
and Yamamoto, Yoshihisa
and Takesue, Hiroki},
title={Large-scale Ising spin network based on degenerate optical parametric oscillators},
journal={Nature Photonics},
year={2016},
month={Jun},
day={01},
volume={10},
number={6},
pages={415-419},
issn={1749-4893},
doi={10.1038/nphoton.2016.68},
url={https://doi.org/10.1038/nphoton.2016.68}
}

@Article{Böhm2019PoorMansCIM,
author={B{\"o}hm, Fabian
and Verschaffelt, Guy
and Van der Sande, Guy},
title={A poor man's coherent Ising machine based on opto-electronic feedback systems for solving optimization problems},
journal={Nature Communications},
year={2019},
month={Aug},
day={08},
volume={10},
number={1},
pages={3538},
issn={2041-1723},
doi={10.1038/s41467-019-11484-3},
url={https://doi.org/10.1038/s41467-019-11484-3}
}

@Article{Johnson2011QuantumAnnealer,
author={Johnson, M. W.
and Amin, M. H. S.
and Gildert, S.
and Lanting, T.
and Hamze, F.
and Dickson, N.
and Harris, R.
and Berkley, A. J.
and Johansson, J.
and Bunyk, P.
and Chapple, E. M.
and Enderud, C.
and Hilton, J. P.
and Karimi, K.
and Ladizinsky, E.
and Ladizinsky, N.
and Oh, T.
and Perminov, I.
and Rich, C.
and Thom, M. C.
and Tolkacheva, E.
and Truncik, C. J. S.
and Uchaikin, S.
and Wang, J.
and Wilson, B.
and Rose, G.},
title={Quantum annealing with manufactured spins},
journal={Nature},
year={2011},
month={May},
day={01},
volume={473},
number={7346},
pages={194-198},
issn={1476-4687},
doi={10.1038/nature10012},
url={https://doi.org/10.1038/nature10012}
}

@article{Mahbood2016-phononicIM,
author = {Imran Mahboob  and Hajime Okamoto  and Hiroshi Yamaguchi },
title = {An electromechanical Ising Hamiltonian},
journal = {Science Advances},
volume = {2},
number = {6},
pages = {e1600236},
year = {2016},
doi = {10.1126/sciadv.1600236},
URL = {https://www.science.org/doi/abs/10.1126/sciadv.1600236},
eprint = {https://www.science.org/doi/pdf/10.1126/sciadv.1600236}}

@Article{Berloff2017PolaritonicIMs,
author={Berloff, Natalia G.
and Silva, Matteo
and Kalinin, Kirill
and Askitopoulos, Alexis
and T{\"o}pfer, Julian D.
and Cilibrizzi, Pasquale
and Langbein, Wolfgang
and Lagoudakis, Pavlos G.},
title={Realizing the classical XY Hamiltonian in polariton simulators},
journal={Nature Materials},
year={2017},
month={Nov},
day={01},
volume={16},
number={11},
pages={1120-1126},
issn={1476-4660},
doi={10.1038/nmat4971},
url={https://doi.org/10.1038/nmat4971}
}

@Article{Kalinin2018-gain-dissipative-systems,
author={Kalinin, Kirill P.
and Berloff, Natalia G.},
title={Global optimization of spin Hamiltonians with gain-dissipative systems},
journal={Scientific Reports},
year={2018},
month={Dec},
day={12},
volume={8},
number={1},
pages={17791},
issn={2045-2322},
doi={10.1038/s41598-018-35416-1},
url={https://doi.org/10.1038/s41598-018-35416-1}
}

@article{Kalinin2018ExternalFields,
  title = {Simulating Ising and $n$-State Planar Potts Models and External Fields with Nonequilibrium Condensates},
  author = {Kalinin, Kirill P. and Berloff, Natalia G.},
  journal = {Phys. Rev. Lett.},
  volume = {121},
  issue = {23},
  pages = {235302},
  numpages = {5},
  year = {2018},
  month = {Dec},
  publisher = {American Physical Society},
  doi = {10.1103/PhysRevLett.121.235302},
  url = {https://link.aps.org/doi/10.1103/PhysRevLett.121.235302}
}

@article{Leleu2019-destabilization-local-minima,
  title = {Destabilization of Local Minima in Analog Spin Systems by Correction of Amplitude Heterogeneity},
  author = {Leleu, Timoth\'ee and Yamamoto, Yoshihisa and McMahon, Peter L. and Aihara, Kazuyuki},
  journal = {Phys. Rev. Lett.},
  volume = {122},
  issue = {4},
  pages = {040607},
  numpages = {6},
  year = {2019},
  month = {Feb},
  publisher = {American Physical Society},
  doi = {10.1103/PhysRevLett.122.040607},
  url = {https://link.aps.org/doi/10.1103/PhysRevLett.122.040607}
}

@Article{Lee2025-noise-augmented-p-bits,
author={Lee, Kyle
and Chowdhury, Shuvro
and Camsari, Kerem Y.},
title={Noise-augmented chaotic Ising machines for combinatorial optimization and sampling},
journal={Communications Physics},
year={2025},
month={Jan},
day={22},
volume={8},
number={1},
pages={35},
issn={2399-3650},
doi={10.1038/s42005-025-01945-1},
url={https://doi.org/10.1038/s42005-025-01945-1}
}

@Article{Bohm2022-noise-injected-IMs,
author={B{\"o}hm, Fabian
and Alonso-Urquijo, Diego
and Verschaffelt, Guy
and Van der Sande, Guy},
title={Noise-injected analog Ising machines enable ultrafast statistical sampling and machine learning},
journal={Nature Communications},
year={2022},
month={Oct},
day={04},
volume={13},
number={1},
pages={5847},
issn={2041-1723},
doi={10.1038/s41467-022-33441-3},
url={https://doi.org/10.1038/s41467-022-33441-3}
}

@article{vadde2026-2048spinbulkacousticwave,
  title={A 2048-spin bulk acoustic wave Ising machine for number partitioning and Sudoku}, 
  author={Venkatesh Vadde and Roman Ovcharov and Victor H. González and Roman Khymyn and Artem Litvinenko and Johan Åkerman},
  year={2026},
  journal={arXiv preprint: [cond-mat.mes-hall] 2607.02112},
  url={https://arxiv.org/abs/2607.02112}, 
}

@article{Cummins2025ManifoldReduction,
  title = {Ising Hamiltonian minimization: Gain-based computing with manifold reduction of soft spins vs quantum annealing},
  author = {Cummins, James S. and Salman, Hayder and Berloff, Natalia G.},
  journal = {Phys. Rev. Res.},
  volume = {7},
  issue = {1},
  pages = {013150},
  numpages = {11},
  year = {2025},
  month = {Feb},
  publisher = {American Physical Society},
  doi = {10.1103/PhysRevResearch.7.013150},
  url = {https://link.aps.org/doi/10.1103/PhysRevResearch.7.013150}
}

@Article{Kalinin2022ComplexityContinuum,
author={Kalinin, Kirill P.
and Berloff, Natalia G.},
title={Computational complexity continuum within Ising formulation of NP problems},
journal={Communications Physics},
year={2022},
month={Jan},
day={12},
volume={5},
number={1},
pages={20},
issn={2399-3650},
doi={10.1038/s42005-021-00792-0},
url={https://doi.org/10.1038/s42005-021-00792-0}
}

@article{litvinenko2023spinwave,
  title={A spinwave Ising machine},
  author={Litvinenko, Artem and Khymyn, Roman and Gonz{\'a}lez, Victor H and Ovcharov, Roman and Awad, Ahmad A and Tyberkevych, Vasyl and Slavin, Andrei and {\AA}kerman, Johan},
  journal={Commun. Phys.},
  volume={6},
  number={1},
  pages={227},
  year={2023},
  publisher={Nature Publishing Group UK London}
}

@article{mohseni2022ising,
  title={Ising machines as hardware solvers of combinatorial optimization problems},
  author={Mohseni, Naeimeh and McMahon, Peter L and Byrnes, Tim},
  journal={Nature Reviews Physics},
  volume={4},
  number={6},
  pages={363--379},
  year={2022},
  publisher={Nature Publishing Group UK London}
}

@ARTICLE{Lucas2014IsingFormulations, 
AUTHOR={Lucas, Andrew},   
TITLE={Ising formulations of many NP problems},      
JOURNAL={Frontiers in Physics},      
VOLUME={2},           
YEAR={2014},      
URL={https://www.frontiersin.org/articles/10.3389/fphy.2014.00005},       
DOI={10.3389/fphy.2014.00005},      
ISSN={2296-424X}
}

@article{Gonzalez2023GlobalZTSWIM,
    author = {González, Victor H. and Litvinenko, Artem and Khymyn, Roman and Åkerman, Johan},
    title = {Global biasing using a hardware-based artificial Zeeman term in spinwave Ising machines},
    journal = {Applied Physics Letters},
    volume = {124},
    number = {9},
    pages = {092409},
    year = {2024},
    month = {02},
    issn = {0003-6951},
    doi = {10.1063/5.0185888},
}

@article{Pierangeli2019-photonic-annealer,
  title = {Large-Scale Photonic Ising Machine by Spatial Light Modulation},
  author = {Pierangeli, D. and Marcucci, G. and Conti, C.},
  journal = {Phys. Rev. Lett.},
  volume = {122},
  issue = {21},
  pages = {213902},
  numpages = {6},
  year = {2019},
  month = {May},
  publisher = {American Physical Society},
  doi = {10.1103/PhysRevLett.122.213902},
  url = {https://link.aps.org/doi/10.1103/PhysRevLett.122.213902}
}

@article{Litvinenko-50spinSAWIM ,
    author = {Artem Litvinenko and Roman Khymyn and Roman Ovcharov and Johan Åkerman},
    title = {A 50-spin surface acoustic wave Ising machine},
    journal =  	{arXiv:2311.06830[cond-mat.mes-hall]} ,
    year = {2023},
    doi={10.48550/arXiv.2311.06830}
}

@Article{Aadit2022-sparsification-fpga,
author={Aadit, Navid Anjum
and Grimaldi, Andrea
and Carpentieri, Mario
and Theogarajan, Luke
and Martinis, John M.
and Finocchio, Giovanni
and Camsari, Kerem Y.},
title={Massively parallel probabilistic computing with sparse Ising machines},
journal={Nat. Electron.},
year={2022},
month={Jul},
day={01},
volume={5},
number={7},
pages={460-468},
issn={2520-1131},
doi={10.1038/s41928-022-00774-2},
url={https://doi.org/10.1038/s41928-022-00774-2}
}

@ARTICLE{Chowdhury2023-full-stack-p-bits,
  author={Chowdhury, Shuvro and Grimaldi, Andrea and Aadit, Navid Anjum and Niazi, Shaila and Mohseni, Masoud and Kanai, Shun and Ohno, Hideo and Fukami, Shunsuke and Theogarajan, Luke and Finocchio, Giovanni and Datta, Supriyo and Camsari, Kerem Y.},
  journal={IEEE Journal on Exploratory Solid-State Computational Devices and Circuits}, 
  title={A Full-Stack View of Probabilistic Computing With p-Bits: Devices, Architectures, and Algorithms}, 
  year={2023},
  volume={9},
  number={1},
  pages={1-11},
  doi={10.1109/JXCDC.2023.3256981}}

@article{Gonzalez2024-spintronic-devices-next-generation-computation,
title = {Spintronic devices as next-generation computation accelerators},
journal = {Curr. Opin. Solid State Mater. Sci.},
volume = {31},
pages = {101173},
year = {2024},
issn = {1359-0286},
doi = {https://doi.org/10.1016/j.cossms.2024.101173},
author = {Victor H. González and Artem Litvinenko and Akash Kumar and Roman Khymyn and Johan \AA{}kerman}
}

@article{ovcharov2024-numerical-model-tm-IM,
  title={A numerical model for time-multiplexed Ising machines based on delay-line oscillators},
  author={Ovcharov, Roman V and Gonz{\'a}lez, Victor H and Litvinenko, Artem and {\AA}kerman, Johan and Khymyn, Roman S},
  journal={arXiv preprint arXiv:2406.07197},
  year={2024},
  url={https://arxiv.org/abs/2406.07197}, 
}

@article{Honjo2021SciAdv100kCIM,
author = {Toshimori Honjo  and Tomohiro Sonobe  and Kensuke Inaba  and Takahiro Inagaki  and Takuya Ikuta  and Yasuhiro Yamada  and Takushi Kazama  and Koji Enbutsu  and Takeshi Umeki  and Ryoichi Kasahara  and Ken-ichi Kawarabayashi  and Hiroki Takesue },
title = {100,000-spin coherent Ising machine},
journal = {Science Advances},
volume = {7},
number = {40},
pages = {eabh0952},
year = {2021},
doi = {10.1126/sciadv.abh0952}}

@article{Inagaki2016Sci2000CIM,
author = {Takahiro Inagaki  and Yoshitaka Haribara  and Koji Igarashi  and Tomohiro Sonobe  and Shuhei Tamate  and Toshimori Honjo  and Alireza Marandi  and Peter L. McMahon  and Takeshi Umeki  and Koji Enbutsu  and Osamu Tadanaga  and Hirokazu Takenouchi  and Kazuyuki Aihara  and Ken-ichi Kawarabayashi  and Kyo Inoue  and Shoko Utsunomiya  and Hiroki Takesue },
title = {A coherent Ising machine for 2000-node optimization problems},
journal = {Science},
volume = {354},
number = {6312},
pages = {603-606},
year = {2016},
doi = {10.1126/science.aah4243}}

@article{McMahon2016Sci100CIM,
author = {Peter L. McMahon  and Alireza Marandi  and Yoshitaka Haribara  and Ryan Hamerly  and Carsten Langrock  and Shuhei Tamate  and Takahiro Inagaki  and Hiroki Takesue  and Shoko Utsunomiya  and Kazuyuki Aihara  and Robert L. Byer  and M. M. Fejer  and Hideo Mabuchi  and Yoshihisa Yamamoto },
title = {A fully programmable 100-spin coherent Ising machine with all-to-all connections},
journal = {Science},
volume = {354},
number = {6312},
pages = {614-617},
year = {2016},
doi = {10.1126/science.aah5178}}

@Article{Inui2022ArtificialZeeman,
author={Inui, Yoshitaka
and Gunathilaka, Mastiyage Don Sudeera Hasaranga
and Kako, Satoshi
and Aonishi, Toru
and Yamamoto, Yoshihisa},
title={Control of amplitude homogeneity in coherent Ising machines with artificial Zeeman terms},
journal={Communications Physics},
year={2022},
month={Jun},
day={15},
volume={5},
number={1},
pages={154},
issn={2399-3650},
doi={10.1038/s42005-022-00927-x},
url={https://doi.org/10.1038/s42005-022-00927-x}
}

@Article{Dutta2021VO2-IMs,
author={Dutta, S.
and Khanna, A.
and Assoa, A. S.
and Paik, H.
and Schlom, D. G.
and Toroczkai, Z.
and Raychowdhury, A.
and Datta, S.},
title={An Ising Hamiltonian solver based on coupled stochastic phase-transition nano-oscillators},
journal={Nature Electronics},
year={2021},
month={Jul},
day={01},
volume={4},
number={7},
pages={502-512},
issn={2520-1131},
doi={10.1038/s41928-021-00616-7},
url={https://doi.org/10.1038/s41928-021-00616-7}
}

@ARTICLE{Zhang2024IMReview,
  author={Zhang, Tingting and Tao, Qichao and Liu, Bailiang and Grimaldi, Andrea and Raimondo, Eleonora and Jiménez, Manuel and Avedillo, María José and Nuñez, Juan and Linares-Barranco, Bernabé and Serrano-Gotarredona, Teresa and Finocchio, Giovanni and Han, Jie},
  journal={IEEE Transactions on Nanotechnology}, 
  title={A Review of Ising Machines Implemented in Conventional and Emerging Technologies}, 
  year={2024},
  volume={23},
  number={},
  pages={704-717},
  doi={10.1109/TNANO.2024.3457533}}

@Article{Perdomo-Ortiz2016CorrectionBiasesQuantumAnnealers,
author={Perdomo-Ortiz, Alejandro
and O'Gorman, Bryan
and Fluegemann, Joseph
and Biswas, Rupak
and Smelyanskiy, Vadim N.},
title={Determination and correction of persistent biases in quantum annealers},
journal={Scientific Reports},
year={2016},
month={Jan},
day={19},
volume={6},
number={1},
pages={18628},
issn={2045-2322},
doi={10.1038/srep18628},
url={https://doi.org/10.1038/srep18628}
}

@ARTICLE{Kasdin1995ColouredNoise,
  author={Kasdin, N.J.},
  journal={Proceedings of the IEEE}, 
  title={Discrete simulation of colored noise and stochastic processes and 1/f/sup /spl alpha// power law noise generation}, 
  year={1995},
  volume={83},
  number={5},
  pages={802-827},
  doi={10.1109/5.381848}}

@article{Takesue2025-CIM40k-dense,
author = {Hiroki Takesue  and Kensuke Inaba  and Toshimori Honjo  and Yasuhiro Yamada  and Takuya Ikuta  and Yuya Yonezu  and Takahiro Inagaki  and Takeshi Umeki  and Ryoichi Kasahara },
title = {Finding independent sets in large-scale graphs with a coherent Ising machine},
journal = {Science Advances},
volume = {11},
number = {7},
pages = {eads7223},
year = {2025},
doi = {10.1126/sciadv.ads7223},
URL = {https://www.science.org/doi/abs/10.1126/sciadv.ads7223},
eprint = {https://www.science.org/doi/pdf/10.1126/sciadv.ads7223}}

@article{takata16bitCIMdelaylines2016,
  title={A 16-bit coherent Ising machine for one-dimensional ring and cubic graph problems},
  author={Takata, Kenta and Marandi, Alireza and Hamerly, Ryan and Haribara, Yoshitaka and Maruo, Daiki and Tamate, Shuhei and Sakaguchi, Hiromasa and Utsunomiya, Shoko and Yamamoto, Yoshihisa},
  journal=sr,
  volume={6},
  number={1},
  pages={1--7},
  year={2016},
  doi = {10.1038/srep34089},
  publisher={Nature Publishing Group}
}

@Inbook{Haribara2016CIMPerformanceEvalDelayLines,
author={Haribara, Yoshitaka
and Utsunomiya, Shoko
and Yamamoto, Yoshihisa},
title={A Coherent Ising Machine for MAX-CUT Problems: Performance Evaluation against Semidefinite Programming and Simulated Annealing},
bookTitle={Principles and Methods of Quantum Information Technologies},
year={2016},
publisher={Springer Japan},
address={Tokyo},
pages={251--262},
doi={10.1007/978-4-431-55756-2\_12}
}

@article{Barahone1988-circuit-design,
author = {Barahona, Francisco and Gr\"{o}tschel, Martin and J\"{u}nger, Michael and Reinelt, Gerhard},
title = {An Application of Combinatorial Optimization to Statistical Physics and Circuit Layout Design},
journal = {Operations Research},
volume = {36},
number = {3},
pages = {493-513},
year = {1988},
doi = {10.1287/opre.36.3.493},

URL = { 
    
        https://doi.org/10.1287/opre.36.3.493
    
    

},
eprint = { 
    
        https://doi.org/10.1287/opre.36.3.493
    
    

}
}

@article{babej2018-protein-folding-quantum-annealer,
  title={Coarse-grained lattice protein folding on a quantum annealer},
    author={Babej, Tom{\'a}{\v{s}} and  Ing, Christopher and Fingerhuth, Mark},
  journal={arXiv preprint arXiv:1811.00713},
  year={2018}
}

@inproceedings{venturelli2016-job-scheduling-quantum-annealer,
  title={Job shop scheduling solver based on quantum annealing},
  author={Venturelli, Davide and Marchand, D and Rojo, Galo},
  booktitle={Proc. of ICAPS-16 Workshop on Constraint Satisfaction Techniques for Planning and Scheduling (COPLAS)},
  pages={25--34},
  year={2016}
}

@article{yamaoka2015-20k-spins,
  title={A 20k-spin Ising chip to solve combinatorial optimization problems with CMOS annealing},
  author={Yamaoka, Masanao and Yoshimura, Chihiro and Hayashi, Masato and Okuyama, Takuya and Aoki, Hidetaka and Mizuno, Hiroyuki},
  journal={IEEE Journal of Solid-State Circuits},
  volume={51},
  number={1},
  pages={303--309},
  year={2015},
  publisher={IEEE}
}

@article{cai2020-memristor-hopfield-networks,
  title={Power-efficient combinatorial optimization using intrinsic noise in memristor Hopfield neural networks},
  author={Cai, Fuxi and Kumar, Suhas and Van Vaerenbergh, Thomas and Sheng, Xia and Liu, Rui and Li, Can and Liu, Zhan and Foltin, Martin and Yu, Shimeng and Xia, Qiangfei and others},
  journal={Nat. Electron.},
  volume={3},
  number={7},
  pages={409--418},
  year={2020},
  publisher={Nature Publishing Group UK London}
}

@article{albash2018-d-wave-computer,
  title={Demonstration of a scaling advantage for a quantum annealer over simulated annealing},
  author={Albash, Tameem and Lidar, Daniel A},
  journal={Physical Review X},
  volume={8},
  number={3},
  pages={031016},
  year={2018},
  publisher={APS}
}

@Article{Borders2019-integer-factorization-mtjs,
author={Borders, William A.
and Pervaiz, Ahmed Z.
and Fukami, Shunsuke
and Camsari, Kerem Y.
and Ohno, Hideo
and Datta, Supriyo},
title={Integer factorization using stochastic magnetic tunnel junctions},
journal={Nature},
year={2019},
month={Sep},
day={01},
volume={573},
number={7774},
pages={390-393},
issn={1476-4687},
doi={10.1038/s41586-019-1557-9},
url={https://doi.org/10.1038/s41586-019-1557-9}
}

@article{Bohm2021,
  author = {B{\"o}hm, Fabian and Vaerenbergh, Thomas Van and Verschaffelt, Guy and Van der Sande, Guy},
  title = {Order-of-magnitude differences in computational performance of analog Ising machines induced by the choice of nonlinearity},
  journal = {Communications Physics},
  year = {2021},
  day = {01},
  volume = {4},
  number = {1},
  pages = {149},
  issn = {2399-3650},
  doi = {10.1038/s42005-021-00655-8},
  url = {https://doi.org/10.1038/s42005-021-00655-8},
}

@article{Barahona1982,
  doi = {10.1088/0305-4470/15/10/028},
  url = {https://dx.doi.org/10.1088/0305-4470/15/10/028},
  year = {1982},
  publisher = {},
  volume = {15},
  number = {10},
  pages = {3241},
  author = {F Barahona},
  title = {On the computational complexity of Ising spin glass models},
  journal = {Journal of Physics A: Mathematical and General},
}

\end{document}